\documentclass[prl,superscriptaddress,twocolumn]{revtex4}%
\usepackage[ansinew]{inputenc}
\usepackage{graphicx}
\usepackage{amsmath}
\usepackage{amsthm}
\usepackage{bm}
\usepackage{layout}
\usepackage{float}
\usepackage{txfonts}
\usepackage{amsfonts}
\usepackage{amssymb}%
\begin{document}

\title{The conditions for quantum violation of macroscopic realism}

\begin{abstract}
Why do we not experience a violation of macroscopic realism in every-day life?
Normally, no violation can be seen either because of decoherence or the
restriction of coarse-grained measurements, transforming the time evolution of
any quantum state into a classical time evolution of a statistical mixture. We
find the sufficient condition for these classical evolutions for spin systems
under coarse-grained measurements. Then we demonstrate that there exist
"non-classical" Hamiltonians whose time evolution cannot be understood
classically, although at every instant of time the quantum spin state appears
as a classical mixture. We suggest that such Hamiltonians are unlikely to be
realized in nature because of their high computational complexity.

\end{abstract}
\date{\today}%

\author{Johannes Kofler}%
%

\affiliation{Fakult\"at f\"ur Physik, Universit\"{a}%
t Wien, Boltzmanngasse 5, 1090 Wien, Austria}%
%

\affiliation
{Institut f\"ur Quantenoptik und Quanteninformation (IQOQI), \"Osterreichische Akademie der Wissenschaften,\\ Boltzmanngasse 3, 1090 Wien, Austria}%
%

\author{{\v C}aslav Brukner}%
%

\affiliation{Fakult\"at f\"ur Physik, Universit\"{a}%
t Wien, Boltzmanngasse 5, 1090 Wien, Austria}%
%

\affiliation
{Institut f\"ur Quantenoptik und Quanteninformation (IQOQI), \"Osterreichische Akademie der Wissenschaften,\\ Boltzmanngasse 3, 1090 Wien, Austria}%
%

\maketitle

The laws of quantum physics are in conflict with a classical world, in
particular with local and macroscopic realism as characterized by the
violation of the Bell~\cite{Bell1964} and
Leggett-Garg~\cite{Legg1985,Legg2002} inequality, respectively. While Bell's
theorem is a well investigated area of research, hardly any analysis has been
undertaken to understand the key ingredients for the violation of macroscopic
realism (macrorealism). Is it the initial state, the Hamiltonian or the
measurement observables which have to be "quantum" to see a deviation from
classical physics?

Macrorealism is defined by the conjunction of three postulates~\cite{Legg2002}%
: "(1) \textit{Macrorealism per se}. A macroscopic object which has available
to it two or more macroscopically distinct states is at any given time in a
definite one of those states. (2) \textit{Non-invasive measurability}. It is
possible in principle to determine which of these states the system is in
without any effect on the state itself or on the subsequent system dynamics.
(3) \textit{Induction}. The properties of ensembles are determined exclusively
by initial conditions (and in particular not by final conditions)." These
assumptions allow to derive Leggett-Garg inequalities.

In this Letter we first show that a violation of the Leggett-Garg inequality
itself is possible for \textit{arbitrary} Hamiltonians given the ability to
distinguish consecutive eigenstates. This is understandable because it is
generally accepted that "microscopically distinct states" do not have
objective existence. For testing macrorealism one needs to apply the
Leggett-Garg definition referring to macroscopically distinct states. In our
every-day life, to experience macrorealism it is usually sufficient to employ
a certain type of decoherence (where the system is \textit{isolated}%
~\footnote{Here we do not consider decoherence models where the system is
continuously monitored by the environment.}\ and only at the times of
measurement the environment makes a pre-measurement on the
apparatus~\cite{Zure1991}) or the restriction of coarse-grained
measurements~\cite{Pere1995,Busc1995,Poul2005,Kofl2006}. While both mechanisms
transform the quantum state at every instance of time into a classical
mixture, we demonstrate that there are "non-classical" Hamiltonians for which
the \textit{time evolution of this mixture cannot be understood classically},
leading to a violation of macrorealism. We find the necessary condition for
non-classical evolutions and illustrate it by the example of a Schrödinger
cat-like state~\cite{Schr1935}. In the last part we argue why such
Hamiltonians are unlikely to be realized.

Consider a physical system and a quantity $A$, which whenever measured is
found to take one of the values $\pm1$ only. Now perform a series of runs
starting from identical initial conditions (at time $t=0$) such that on the
first set of runs $A$ is measured only at times $t_{1}$ and $t_{2}$, only at
$t_{2}$ and $t_{3}$ on the second, and at $t_{1}$ and $t_{3}$ on the third
$(0\!\leq\!t_{1}\!<\!t_{2}\!<\!t_{3})$. Introducing temporal correlation
functions $C_{ij}\equiv\langle A(t_{i})\,A(t_{j})\rangle$, any macrorealistic
theory predicts Leggett-Garg inequalities, for instance of the Wigner
type~\cite{Wign1970}:%
\begin{equation}
K\equiv C_{12}+C_{23}-C_{13}\leq1\,. \label{eq Herbert}%
\end{equation}

\textit{Any non-trivial (time-independent) Hamiltonian} $\hat{H}$ leads to a
violation of this inequality. We extend the approach of Peres in
Ref.~\cite{Pere1995} and look at the "survival probability" of the system's
initial state at time $t=0$. This state be denoted as $|\psi(0)\rangle
\equiv|\psi_{0}\rangle$ (which must not be an energy eigenstate) and, without
measurements, it evolves to $|\psi(t)\rangle=\exp(-$i$\hat{H}t/\hbar
)\,|\psi_{0}\rangle$ according to the Schrödinger equation. Our dichotomic
observable is $\hat{A}\equiv2\,|\psi_{0}\rangle\langle\psi_{0}|-\openone$,
i.e.\ we ask whether the system is (still) in the state $|\psi_{0}\rangle$
(outcome '$+$' $\equiv+1$) or not (outcome '$-$' $\equiv-1$). The temporal
correlations $C_{ij}$ can be written as $C_{ij}=p_{i+}\,q_{j+|i+}%
+p_{i-}\,q_{j-|i-}-p_{i+}\,q_{j-|i+}-p_{i-}\,q_{j+|i-}$, where $p_{i+}$
($p_{i-}$) is the probability for measuring '$+$' ('$-$') at $t_{i}$ and
$q_{jl|ik}$ is the probability for measuring $l$ at $t_{j}$ given that $k$ was
measured at $t_{i}$ ($k,l=+,-$). For simplicity we choose $t_{1}=0$ and
equidistant times $\Delta t\equiv t_{2}-t_{1}=t_{3}-t_{2}$. Then the
correlation $C_{12}$ is given by $C_{12}=2p(\Delta t)-1$, where $p(t)\equiv
|\langle\psi_{0}|\psi(t)\rangle|^{2}$ is the (survival) probability to find
$|\psi_{0}\rangle$ given the state $|\psi(t)\rangle$. Analogously, we find
$C_{13}=2p(2\Delta t)-1$ and $C_{23}$. Plugging everything into
(\ref{eq Herbert}), one ends up with%
\begin{equation}
K=4\,p(\Delta t)\sqrt{p(2\Delta t)}\,\cos\gamma-4\,p(2\Delta t)+1\leq1\,,
\label{eq Herbert P}%
\end{equation}
where $\gamma\equiv2\alpha-\beta$ and $\alpha$ and $\beta$ are the phases in
$\langle\psi_{0}|\psi(t_{2})\rangle=\!\sqrt{p(\Delta t)}\,$e$^{\text{i}\alpha
}$ and $\langle\psi_{0}|\psi(t_{3})\rangle=\!\sqrt{p(2\Delta t)}%
\,$e$^{\text{i}\beta}$.

Now, independent of the system's dimension, it is sufficient to consider as
initial state a superposition of only two energy eigenstates $|u_{1}\rangle$
and $|u_{2}\rangle$ with energy eigenvalues $E_{1}$ and $E_{2}$, respectively:
$|\psi_{0}\rangle=(|u_{1}\rangle\!+\!|u_{2}\rangle)/\!\sqrt{2}$.
Ineq.\ (\ref{eq Herbert P}) becomes $K=2\cos(\tfrac{\Delta E\Delta t}{\hbar
})-\cos(\tfrac{2\Delta E\Delta t}{\hbar})\leq1$, with $\Delta E\equiv
E_{2}-E_{1}$ the energy difference of the two levels, and a violation is
always possible. The left hand side reaches $K=1.5$ for $\Delta t=\frac
{\pi\hbar}{3\Delta E}$ and $\Delta t=\frac{5\pi\hbar}{3\Delta E}$ and in
$\frac{2\pi\hbar}{\Delta E}$ periods thereof.

Why then do we not see a violation of the Leggett-Garg inequality in everyday
life? The usual answer is that this is either due to decoherence or due to the
fact that the resolution of our everyday measurements is not sharp, making it
impossible to project onto individual states and hence making it impossible to
see the above demonstrated violation that is always present for microstates.

For testing macrorealism---i.e.\ testing the Leggett-Garg inequality under the
restriction of \textit{coarse-grained measurements}---we consider a spin-$j$
system (with $j\!\gg\!1$) as a model example. Any spin-$j$ state can be
written in the quasi-diagonal form $\hat{\rho}=%
{\textstyle\iint}
P(\Omega)\,|\Omega\rangle\langle\Omega|\,$d$^{2}\Omega$ with d$^{2}\Omega$ the
solid angle element and $P$ a normalized and \textit{not necessarily positive}
real function~\cite{Arec1972}. The\textit{ }spin coherent states\textit{
}$|\Omega\rangle\equiv|\vartheta,\varphi\rangle$, with $\vartheta$ and
$\varphi$ the polar and azimuthal angle, are the eigenstates with maximal
eigenvalue of a spin operator pointing into the direction $\Omega
\equiv(\vartheta,\varphi)$~\cite{Radc1971}: $\hat{\mathbf{J}}_{\Omega
}\left\vert \Omega\right\rangle =j\left\vert \Omega\right\rangle $ in units
where $\hbar=1$. In coarse-grained measurements our resolution is not able to
resolve individual eigenvalues $m$ of a spin component, say the $z$-component
$\hat{J}_{z}$, but bunches together $\Delta m$ \textit{neighboring}%
~\footnote{The term "neighboring"\ only makes sense in a classical context,
treating those eigenvalues as close which correspond to close outcomes in real
phase space. In Hilbert space $|m\rangle$ and $|m\!+\!1\rangle$ are as
orthogonal as $|m\rangle$ and $|m\!+\!10^{10}\rangle$.} outcomes into "slots"
$\bar{m}$, where the measurement coarseness is much larger than the intrinsic
uncertainty of coherent states, i.e. $\Delta m\!\gg\!\!\sqrt{j}$%
~\cite{Kofl2006}.

The question arises whether it is problematic to use coarse-grained
\textit{von Neumann measurements} of the form $%
{\textstyle\sum\nolimits_{m\in\{\bar{m}\}}}
\!\left\vert m\right\rangle \!\left\langle m\right\vert $, where $\left\vert
m\right\rangle $ are the $\hat{J}_{z}$ eigenstates, as "classical
measurements". In contrast to the \textit{positive operator value measure}
(POVM), they have sharp edges and could violate the Leggett-Garg inequality by
distinguishing with certainty between microstates at two sides of a slot
border. Therefore, we model our coarse-grained $\hat{J}_{z}$ measurements as
belonging to a (spin coherent state) POVM, where the element corresponding to
the outcome $\bar{m}$ is represented by%
\begin{equation}
\hat{P}_{\bar{m}}\equiv\tfrac{2j+1}{4\pi}\,%
{\textstyle\iint\nolimits_{\Omega_{\bar{m}}}}
|\Omega\rangle\langle\Omega|\,\text{d}^{2}\Omega\,.
\end{equation}
Here, $\Omega_{\bar{m}}$ is the angular region of polar angular size
$\Delta\Theta_{\bar{m}}\sim\Delta m/j\gg1/\!\sqrt{j}$ whose projection onto
the $z$ axis corresponds to the slot $\bar{m}$. As the $\Omega_{\bar{m}}$ are
mutually disjoint and form a partition of the whole angular region, we have $%
{\textstyle\sum\nolimits_{\bar{m}}}
\hat{P}_{\bar{m}}=\openone$. The POVM elements are overlapping at the slot
borders over the angular size $\sim\!1/\!\sqrt{j}$ which is small compared to
the angular slot size $\Delta\Theta_{\bar{m}}$. In the basis of $\hat{J}_{z}$
eigenstates $\hat{P}_{\bar{m}}=%
{\textstyle\sum\nolimits_{k=-j}^{j}}
\tfrac{2j+1}{4\pi}%
{\textstyle\iint\nolimits_{\Omega_{\bar{m}}}}
|\langle k|\Omega\rangle|^{2}\,$d$^{2}\Omega\,|k\rangle\langle k|$ is diagonal
where $|\langle k|\Omega\rangle|^{2}=\left(  \!%
\genfrac{}{}{0pt}{1}{2j}{j+k}%
\!\right)  \cos^{2(j+k)}\!\tfrac{\vartheta}{2}\sin^{2(j-k)}\!\tfrac{\vartheta
}{2}$.

The probability for getting the particular outcome $\bar{m}$ is given by
$w_{\bar{m}}=\;$Tr$[\hat{\rho}\hat{P}_{\bar{m}}]=\tfrac{2j+1}{4\pi}%
{\textstyle\iint}
\langle\Omega|\hat{\rho}\hat{P}_{\bar{m}}|\Omega\rangle\,$d$^{2}\Omega$. This
probability can (exactly) be computed via integration of an \textit{ensemble
of classical spins} over the region $\Omega_{\bar{m}}$, i.e.$\ w_{\bar{m}}=%
{\textstyle\iint\nolimits_{\Omega_{\bar{m}}}}
Q(\Omega)\,$d$^{2}\Omega$, with a \textit{positive probability distribution}
(the well-know $Q$-function~\cite{Agar1981}):%
\begin{equation}
Q(\Omega)\equiv\tfrac{2j+1}{4\pi}\,\langle\Omega|\hat{\rho}|\Omega\rangle\,.
\label{eq Q}%
\end{equation}
That shows that under fuzzy measurements any quantum state allows a classical
description (i.e.\ a hidden variable model). This is \textit{macrorealism per
se}.

Upon a coarse-grained measurement with outcome $\bar{m}$, the state $\hat
{\rho}$ is reduced to $\hat{\rho}_{\bar{m}}=\hat{M}_{\bar{m}}\,\hat{\rho
}\,\hat{M}_{\bar{m}}/w_{\bar{m}}$ where we have chosen a particular
(optimal~\cite{Pere1995}) implementation of the POVM with the Hermitean Kraus
operators $\hat{M}_{\bar{m}}=\hat{M}_{\bar{m}}^{\dag}=%
{\textstyle\sum\nolimits_{k=-j}^{j}}
\left(  \tfrac{2j+1}{4\pi}%
{\textstyle\iint\nolimits_{\Omega_{\bar{m}}}}
|\langle k|\Omega\rangle|^{2}\,\text{d}^{2}\Omega\right)  \!^{1/2}%
\,|k\rangle\langle k|$\ satisfying $\hat{M}_{\bar{m}}^{2}=\hat{P}_{\bar{m}}$.
We note that, independently of the implementation, the $\hat{P}_{\bar{m}}$
(and the Kraus operators) behave almost as projectors for all states
$|\Omega\rangle$ except for those near a slot border. In a proper classical
limit ($\!\sqrt{j}/\Delta m\rightarrow0$) the relative weight of these
$\Omega$ compared to the whole sphere surface becomes \textit{vanishingly
small}. The $Q$-distribution before the measurement is the (weighted)
\textit{mixture} of the $Q$-distributions $Q_{\bar{m}}(\Omega)=\tfrac
{2j+1}{4\pi}\,\langle\Omega|\hat{\rho}_{\bar{m}}|\Omega\rangle$ of the
possible reduced states:%
\begin{equation}
Q(\Omega)\approx%
{\displaystyle\sum\nolimits_{\bar{m}}}
w_{\bar{m}}\,Q_{\bar{m}}(\Omega)\,. \label{eq Q1}%
\end{equation}
The approximate sign "$\approx$" reflects that, depending on the density
matrix $\hat{\rho}\equiv%
{\textstyle\sum\nolimits_{n}}
{\textstyle\sum\nolimits_{n^{\prime}}}
c_{nn^{\prime}}|n\rangle\langle n^{\prime}|$, this relationship may only
approximately hold for the set of those $\Omega\equiv(\vartheta,\varphi)$ near
a slot border. In detail eq.~(\ref{eq Q1}) reads: $\tfrac{2j+1}{4\pi}%
{\textstyle\sum\nolimits_{n}}
{\textstyle\sum\nolimits_{n^{\prime}}}
c_{nn^{\prime}}\langle\Omega|n\rangle\langle n^{\prime}|\Omega\rangle
\approx\tfrac{2j+1}{4\pi}%
{\textstyle\sum\nolimits_{n}}
{\textstyle\sum\nolimits_{n^{\prime}}}
c_{nn^{\prime}}[%
{\textstyle\sum\nolimits_{\bar{m}}}
\!\sqrt{g_{\bar{m}}(n)\,g_{\bar{m}}(n^{\prime})}]\langle\Omega|n\rangle\langle
n^{\prime}|\Omega\rangle$ with $g_{\bar{m}}(k)\equiv\tfrac{2j+1}{4\pi}%
{\textstyle\iint\nolimits_{\Omega_{\bar{m}}}}
|\langle k|\Omega\rangle|^{2}\,$d$^{2}\Omega$ which is smaller or equal to 1.
Deviations only occur if $n$, $n^{\prime}$ ($n^{\prime}\!\neq\!n$) and
$j\cos\vartheta$ are all within a distance of order $\!\sqrt{j}$ to each other
and to a slot border~\footnote{If they are not close to each other, the
quantity $\langle\Omega|n\rangle\langle n^{\prime}|\Omega\rangle$ is
exponentially small and suppression by the factor $%
{\textstyle\sum\nolimits_{\bar{m}}}
\!\sqrt{g_{\bar{m}}(n)\,g_{\bar{m}}(n^{\prime})}$ is not important. If $n$,
$n^{\prime}$ are well within a slot $%
{\textstyle\sum\nolimits_{\bar{m}}}
\!\sqrt{g_{\bar{m}}(n)\,g_{\bar{m}}(n^{\prime})}$ is almost identical to $1$%
.}. Even in the case of a spin coherent state exactly on a slot border, the
overlap between the left and right hand side of eq.~(\ref{eq Q1}) is
$\approx\!0.997$ (independent of $j$), where the overlap of two probability
distributions $f$ and $g$ is defined as $%
{\textstyle\iint}
\!\sqrt{f(\Omega)\,g(\Omega)}\,$d$^{2}\Omega\in\lbrack0,1]$. Eq.\ (\ref{eq Q1}%
) thus shows that a fuzzy measurement can be understood classically as
reducing the previous ignorance about predetermined properties of the spin
system~\cite{Kofl2006}.

Consider the initial distribution of classical spins, $Q(\Omega,t_{0})$,
corresponding to an initial quantum state $\hat{\rho}(t_{0})$. We first
compute the $Q$-distribution of the state $\hat{\rho}(t_{j})$ for an
undisturbed evolution without measurement until some time $t_{j}$,
$Q(\Omega,t_{j})=\tfrac{2j+1}{4\pi}\,\langle\Omega|\hat{\rho}(t_{j}%
)|\Omega\rangle$. This has to be compared with the \textit{mixture} of all
possible reduced distributions upon measurement at a time $t_{i}$
($t_{0}\!\leq\!t_{i}\!<\!t_{j}$) with outcomes $\bar{m}$ which evolved to
$t_{j}$, denoted as $Q_{\bar{m},t_{i}}(\Omega,t_{j})=\tfrac{2j+1}{4\pi
}\,\langle\Omega|\hat{U}_{t_{j}-t_{i}}\hat{M}_{\bar{m}}\,\hat{\rho}%
(t_{i})\,\hat{M}_{\bar{m}}\hat{U}_{t_{j}-t_{i}}^{\dag}|\Omega\rangle
/w_{\bar{m},t_{i}}$ with $w_{\bar{m},t_{i}}\equiv\;$Tr$[\hat{\rho}(t_{i}%
)\hat{P}_{\bar{m}}]$ and $\hat{U}_{t}\equiv\exp(-$i$\hat{H}t)$ the time
evolution operator. The system evolves macrorealistically if these two
quantities coincide for all $t_{i}$ and $t_{j}$,%
\begin{equation}
Q(\Omega,t_{j})\approx%
{\displaystyle\sum\nolimits_{\bar{m}}}
w_{\bar{m},t_{i}}\,Q_{\bar{m},t_{i}}(\Omega,t_{j})\,. \label{eq Q cond}%
\end{equation}
This is \textit{non-invasive measurability} together with \textit{induction}.

In a dichotomic scenario the outcomes '$+$' and '$-$' correspond to finding
the spin system in one out of only two slots $\bar{m}=\pm1$. This is
represented by a measurement of two complementary regions $\Omega_{+}$ and
$\Omega_{-}$ (for instance the northern and southern hemisphere in a "which
hemisphere" measurement). Then, e.g., the probability for measuring '$-$' at
$t_{3}$ if '$+$' was measured at $t_{1}$ is given by $q_{3-|1+}=%
{\textstyle\iint\nolimits_{\Omega_{-}}}
Q_{+,t_{1}}(\Omega,t_{3})\,$d$^{2}\Omega$ with $Q_{+,t_{1}}(\Omega,t_{3})$ the
$Q$-distribution of the state which was reduced at $t_{1}$ with outcome '$+$'
and evolved to $t_{3}$. If condition~(\ref{eq Q cond}) is satisfied, it
implies that the probabilities can be decomposed into "classical paths". This
means that, e.g., $q_{3-|1+}$ is just the sum of the two possible paths via
'$+$' and '$-$' at $t_{2}$: $q_{3-|1+}=q_{2+|1+}\,q_{3-|2+,1+}+q_{2-|1+}%
\,q_{3-|2-,1+}$, where $q_{3-|2\pm,1+}$ denotes the probability to measure
'$-$' at $t_{3}$ given that '$+$' was measured at $t_{1}$ and '$\pm$' at
$t_{2}$. Thus, eq.~(\ref{eq Q cond}) allows to derive Leggett-Garg
inequalities such as (\ref{eq Herbert}).

We can now establish the \textit{sufficient condition for macrorealism} that
holds even for isolated systems, namely%
\begin{equation}
\hat{P}_{\bar{m}}\,\hat{U}_{t}\,|\Omega\rangle\approx\left\{
\begin{array}
[c]{ll}%
\hat{U}_{t}\,|\Omega\rangle & \text{for one }\bar{m},\\
\mathbf{0} & \text{for all the others,}%
\end{array}
\right.  \label{eq U cond}%
\end{equation}
for all $t$ and $\Omega$, allowing deviations at slot borders. This means that
$\hat{U}_{t}$ does not produce superpositions of macroscopically distinct
states and therefore $\hat{P}_{\bar{m}}$, and hence $\hat{M}_{\bar{m}}$, quasi
behave as projectors. Eq.~(\ref{eq U cond}) implies $\langle\Omega|\hat
{U}_{t_{j}-t_{i}}\,\hat{\rho}(t_{i})\,\hat{U}_{t_{j}-t_{i}}^{\dag}%
|\Omega\rangle\approx%
{\textstyle\sum\nolimits_{\bar{m}}}
\langle\Omega|\hat{U}_{t_{j}-t_{i}}\hat{M}_{\bar{m}}\,\hat{\rho}(t_{i}%
)\,\hat{M}_{\bar{m}}\hat{U}_{t_{j}-t_{i}}^{\dag}|\Omega\rangle$ which directly
leads to eq.~(\ref{eq Q cond}). Thus, eq.~(\ref{eq U cond}) $\rightarrow$
eq.~(\ref{eq Q cond}) $\rightarrow$ macrorealism.

We denote those Hamiltonians for which eq.~(\ref{eq U cond}) is satisfied
under coarse-grained measurements as \textit{classical}. An example is the
rotation, say around $x$, $\hat{H}=\omega\hat{J}_{x}$, with $\hat{J}_{x}$ the
spin $x$-component and $\omega$ the angular precession frequency, which
satisfies eq.~(\ref{eq U cond}) and moreover allows a Newtonian description of
the time evolution~\cite{Kofl2006}. But there is no \textit{a priori} reason
why all Hamiltonians should satisfy eq.~(\ref{eq U cond}). Can one find
\textit{non-classical} Hamiltonians violating macrorealism despite
coarse-grained measurements? The necessary condition for this is that the
Hamiltonian builds up coherences between states belonging to different slots.
One explicit (extreme) example is%
\begin{equation}
\hat{H}=\text{i\thinspace}\omega\left(  \left\vert -j\right\rangle
\!\left\langle +j\right\vert -\left\vert +j\right\rangle \!\left\langle
-j\right\vert \right)  , \label{eq Schroe}%
\end{equation}
which, given the special initial state $|\Psi(0)\rangle=\left\vert
+j\right\rangle $, produces a time-dependent Schrödinger cat-like
superposition of two \textit{distant} (orthogonal) spin-$j$ coherent states
$\left\vert +j\right\rangle $ and $\left\vert -j\right\rangle $:%
\begin{equation}
|\Psi(t)\rangle=\cos(\omega t)\left\vert +j\right\rangle +\sin(\omega
t)\left\vert -j\right\rangle . \label{eq psi}%
\end{equation}
Under fuzzy measurements or pre-measurement decoherence~\cite{Zure1991}, the
state~(\ref{eq psi}) appears like a statistical mixture at every instance of
time:%
\begin{equation}
\hat{\rho}_{\text{mix}}(t)=\cos^{2}(\omega t)\left\vert +j\right\rangle
\!\left\langle +j\right\vert +\sin^{2}(\omega t)\left\vert -j\right\rangle
\!\left\langle -j\right\vert . \label{eq mix}%
\end{equation}
While the two states $\hat{\rho}_{\text{sup}}(t)\equiv|\Psi(t)\rangle
\langle\Psi(t)|$ and $\hat{\rho}_{\text{mix}}(t)$, having different
$P$-functions (Fig.~1), can be distinguished by sharp measurements, they are
equivalent on the coarse-grained level. The $Q$-distributions, $Q_{\text{sup}%
}$ for $\hat{\rho}_{\text{sup}}(t)$ and $Q_{\text{mix}}$ for $\hat{\rho
}_{\text{mix}}(t)$, are given by eq.~(\ref{eq Q}). The coherence terms
stemming from $\hat{\rho}_{\text{sup}}(t)$ are of the form $\langle
\Omega\left\vert +j\right\rangle \!\left\langle -j\right\vert \Omega\rangle$
and vanish exponentially fast with the spin length $j$ for all $\Omega$. For
$j\gg1$ the $Q$-distributions are practically identical, i.e.\ $Q_{\text{sup}%
}(\Omega,t)\approx Q_{\text{mix}}(\Omega,t)=\tfrac{2j+1}{4\pi}[\cos^{2}(\omega
t)\cos^{4j}(\tfrac{\Theta_{1}}{2})+\sin^{2}(\omega t)\cos^{4j}(\tfrac
{\Theta_{2}}{2})]$, where $\Theta_{1}=\vartheta$ ($\Theta_{2}=\pi-\vartheta$)
is the angle between $\Omega\equiv(\vartheta,\varphi)$ and $+z$ ($-z$). The
$P$ and $Q$-functions of $\hat{\rho}_{\text{sup}}$ and $\hat{\rho}%
_{\text{mix}}$ at $t=\frac{\pi}{4\omega}$ are shown in Fig.~1 for a certain
choice of parameters~\footnote{Ref.~\cite{Agar1981} shows how to compute the
$P$-functions.}. Using a dichotomic "which hemisphere" measurement, the
temporal correlation function reads $C_{ij}\approx\cos[\omega(t_{j}%
\!-\!t_{i})]$. The system effectively behaves as a spin-$\tfrac{1}{2}$
particle and violates macrorealism. In agreement, eqs.~(\ref{eq Q cond})
and~(\ref{eq U cond}) are not fulfilled. To get macrorealism one would have to
coarse-grain always those states which are \textit{connected} by the
Hamiltonian and not necessarily in real space. In the present case it is (at
least) the outcomes '$+j$' and '$-j$' which have to be coarse-grained into one
and the same slot, which is of course highly counter-intuitive. Such a
coarse-graining would lead to a different kind of macrorealistic physics than
the classical laws we know, bringing systems through space and time
continuously.\begin{figure}[t]
\begin{center}
\includegraphics[width=.45\textwidth]{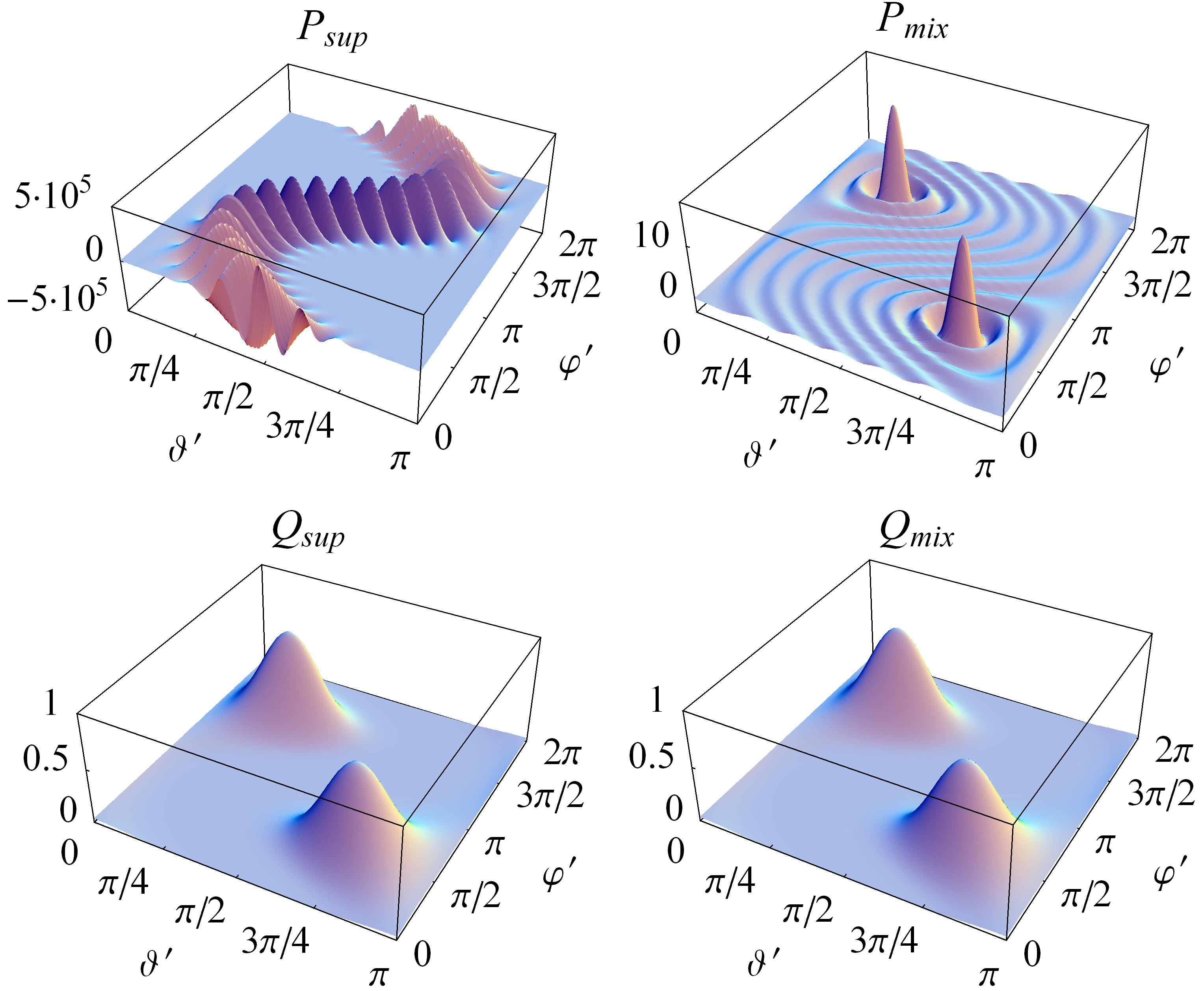}
\end{center}
\par
\vspace{-0.33cm}\caption{(Color online.) Top left: The wildly oscillating
$P$-function $P_{\text{sup}}$ at time $t=\frac{\pi}{4\omega}$ of the
equal-weight superposition (\ref{eq psi}) of two opposite spin coherent states
$\left\vert +j\right\rangle $ and $\left\vert -j\right\rangle $ for spin
length $j=10$, plotted in a rotated coordinate system in which $\left\vert
+j\right\rangle =|\frac{\pi}{4},\frac{3\pi}{2}\rangle$. Top right: The
$P$-function $P_{\text{mix}}$ of the corresponding statistical mixture
(\ref{eq mix}). Bottom: In every-day life the angular measurement resolution
is much weaker than $1/\!\sqrt{j}$. Then we cannot distinguish anymore between
the superposition state and the classical mixture, as both lead to the same
(positive) $Q$-distribution $Q_{\text{sup}}\approx Q_{\text{mix}}$.
Nevertheless, the time evolution of such a mixture can violate macrorealism
even under classical (coarse-grained) measurements.}%
\end{figure}

Finally, we suggest a possible reason why non-classical evolutions might be
unlikely to be realized by nature: Such evolutions either require Hamiltonians
with many-particle interactions or a specific sequence of a large number of
computational steps if only few-particle interactions are used ("high
computational complexity"). Both cases intuitively seem to be of very low
probability to happen spontaneously. Consider our spin-$j$ as a macroscopic
ensemble of $N$ spin-$\tfrac{1}{2}$ particles (i.e.\ qubits) such as, e.g.,
any magnetic material is constituted by many individual microscopic spins. For
violating macrorealism it is necessary to build up superpositions of two
macroscopically distinct coherent states~\footnote{For large $j$ their angular
separation $\Delta\theta$ can be very small and only has to obey the
coarse-graining condition $\Delta\theta\!\gg1/\!\sqrt{j}$. This guarantees
quasi-orthogonality as their (modulus square) overlap is $\cos^{4j}%
(\Delta\theta/2)\sim\exp(-j\Delta\theta^{2})$.}. Without loss of generality we
consider again the particular Hamiltonian~(\ref{eq Schroe}). If $|1\rangle$
and $|0\rangle$ denote the individual qubit states 'up' and 'down' along $z$,
then $|11...1\rangle$ and $|00...0\rangle$ form the total coherent states
$\left\vert +j\right\rangle $ and $\left\vert -j\right\rangle $. The
Hamiltonian represents $N$-particle interactions of the form $\hat{H}%
=\tfrac{\text{i}}{2}\,(\hat{\sigma}_{-}^{\otimes N}\!-\!\hat{\sigma}%
_{+}^{\otimes N})$ where $\hat{\sigma}_{\pm}\equiv\hat{\sigma}_{x}\pm
\,$i$\,\hat{\sigma}_{y}$ with $\hat{\sigma}_{x}$ and $\hat{\sigma}_{y}$ the
Pauli operators. As an alternative one can simulate the evolution governed by
this many-body interaction by means of a series of (in nature typically
appearing) few-qubit interactions (gates), using the methods of quantum
computation science~\cite{Niel2000}. The task is to simulate%
\begin{equation}
|11...1\rangle\;\rightarrow\;\cos(\omega t)\,|11...1\rangle+\sin(\omega
t)\,|00...0\rangle\,. \label{eq unitary2}%
\end{equation}
Assuming sequential qubit interactions, we start from the state
$|11...1\rangle$ and rotate the first qubit '1' by a small angle $\omega\Delta
t$: $|1\rangle_{1}\rightarrow\cos(\omega\Delta t)\,|1\rangle_{1}+\sin
(\omega\Delta t)\,|0\rangle_{1}$. Then we perform a controlled-not (c-not)
gate between this qubit '1' and qubit '2' such that $|x\rangle_{1}%
|y\rangle_{2}\rightarrow|x\rangle_{1}|x\!\oplus\!y\rangle_{2}$ ($x,y=0,1$).
Afterwards c-nots between qubits are performed such that all other qubits are
reached (Fig.~2). This procedure brings us to the state at time $\Delta t$:
$|11...1\rangle\rightarrow\cos(\omega\Delta t)\,|11...1\rangle+\sin
(\omega\Delta t)\,|00...0\rangle$. To simulate the next time interval $\Delta
t$, we have to undo all the c-nots, rotate the first qubit again by
$\omega\Delta t$, and make all the c-nots again, leading to the correct state
at time $2\Delta t$. With this procedure we get a sequence of states,
simulating the evolution (\ref{eq unitary2}). One needs $O(N)$ computational
steps per interval $\Delta t$~\footnote{This is known to be optimal in the
case where qubits can interact only sequentially~\cite{Brav2006}. Relaxing
this condition and permitting simultaneous two-qubit interactions, allows to
decrease the number of sequential steps but does not change the total number
$O(N)$ of necessary gates per interval.}. Note for comparison, however, that
the rotation (say around $x$), $\hat{H}=\tfrac{\omega}{2}%
{\textstyle\sum\nolimits_{k=1}^{N}}
\hat{\sigma}_{x}^{(k)}$ with $k$ labeling the qubits, does not require
interations between qubits. Moreover, the simulation of an interval $\Delta t$
of a spin rotation of the whole chain, i.e.\ $|111...\rangle\rightarrow
\lbrack\cos(\omega\Delta t)\,|1\rangle+\sin(\omega\Delta t)\,|0\rangle
]^{\otimes N}$, can be achieved in a \textit{single global transformation} on
all qubits simultaneously. While both evolutions are rotations in Hilbert
space (and require only polynomial resources), the simulation of the
"non-classical" cosine-law between states that are distant in real space
is---for macroscopically large $N$---computationally much more complex than
the "classical" rotation in real space~\footnote{One should, however, mention
the possibility that an \textit{external} field may produce an effectively
simple non-classical Hamiltonian for \textit{N} qubits where the field
interacts with the collective modes $\left\vert +j\right\rangle
=|11...1\rangle$ and $\left\vert -j\right\rangle =|00...0\rangle$%
.}.\begin{figure}[t]
\begin{center}
\includegraphics[width=6cm]{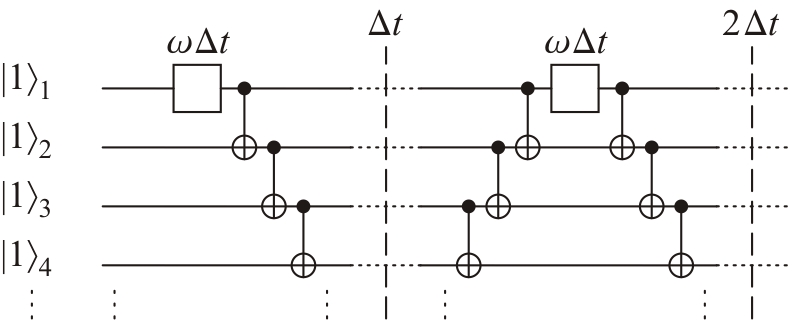}
\end{center}
\par
\vspace{-0.33cm}\caption{To simulate the time evolution (\ref{eq unitary2}) of
a chain of $N$ qubits one needs $O(N)$ computational steps per time interval
$\Delta t$.}%
\end{figure}

\textit{Conclusion}.---Under sharp measurements any non-trivial Hamiltonian is
in conflict with a classical time evolution. Under coarse-grained measurements
any quantum spin state appears as a statistical mixture of spins at every
instance of time. For classical Hamiltonians these mixtures have a classical
time evolution and satisfy macrorealism. Non-classical Hamiltonians build up
quantum coherences between macroscopically distinct states, leading to a
violation of macrorealism. Such Hamiltonians, however, require interactions
between a large number of particles or are computationally much more complex
than classical Hamiltonians, which might be the reason why they are unlikely
to appear in nature.

We thank A.~J.~Leggett, T.~Paterek, F.~Verstraete, A.~Zeilinger, and an
anonymous referee for helpful remarks. This work was supported by the Austrian
Science Foundation FWF (Project No.\ P19570-N16), the European Commission
through Project QAP (No.\ 015846), and the FWF Doctoral Program CoQuS.
J.~K.~is recipient of a DOC fellowship of the Austrian Academy of Sciences.


\begin{thebibliography}{99}                                                                                               %


\bibitem {Bell1964}J. S. Bell, Physics (New York) \textbf{1}, 195 (1964).

\bibitem {Legg1985}A. J. Leggett and A. Garg, Phys. Rev. Lett. \textbf{54},
857 (1985).

\bibitem {Legg2002}A. J. Leggett, J. Phys.: Cond. Mat. \textbf{14}, R415 (2002).

\bibitem {Zure1991}W. H. Zurek, Phys. Today \textbf{44}, 36 (1991); W. H.
Zurek, Rev. Mod. Phys. \textbf{75}, 715 (2003).

\bibitem {Pere1995}A. Peres, \textit{Quantum Theory:~Concepts and Methods}
(Kluwer Academic Publishers, 1995).

\bibitem {Busc1995}P. Busch \textit{et al}., \textit{Operational Quantum
Physics} (Springer, 1995).

\bibitem {Poul2005}D. Poulin, Phys. Rev. A \textbf{71}, 022102 (2005).

\bibitem {Kofl2006}J. Kofler and \v{C}. Brukner, Phys. Rev. Lett. \textbf{99},
180403 (2007).

\bibitem {Schr1935}E. Schrödinger, Die Naturwissenschaften \textbf{48}, 807 (1935).

\bibitem {Wign1970}E. P. Wigner, Am. J. Phys. \textbf{38}, 1005 (1970).

\bibitem {Arec1972}F. T. Arecchi \textit{et al}., Phys. Rev. A \textbf{6},
2211 (1972).

\bibitem {Radc1971}J. M. Radcliffe, J. Phys. A: Gen. Phys. \textbf{4}, 313
(1971); P. W. Atkins and J. C. Dobson, Proc. R. Soc. A \textbf{321}, 321 (1971).

\bibitem {Agar1981}G. S. Agarwal, Phys. Rev. A \textbf{24}, 2889 (1981); G. S.
Agarwal, Phys. Rev. A \textbf{47}, 4608 (1993).

\bibitem {Niel2000}M. A. Nielsen and I. L. Chuang, \textit{Quantum Computation
and Quantum Information} (Cambridge University Press, 2000).

\bibitem {Brav2006}S. Bravyi \textit{et al}., Phys. Rev. Lett. \textbf{97},
050401 (2006).
\end{thebibliography}
\end{document}